\documentclass[aps,prl,twocolumn,showpacs,showkeys,preprintnumbers,amsmath,amssymb,superscriptaddress]{revtex4}

\usepackage{graphicx}

\begin{document}


\def\Tc{$T_{c}$ }
\def\Hc2{$H$$_{c2}$}
\def\ncow{Na$_{0.31}$CoO$_{2}\cdot$1.3H$_{2}$O }

\title{Superconductivity in a layered cobalt oxyhydrate Na$_{0.31}$CoO$_{2}\cdot$1.3H$_{2}$O}
\author{Guanghan \surname{Cao}}
\thanks{Author to whom correspondence should be addressed}
\email{ghcao@zju.edu.cn} \affiliation{Department of Physics,
Zhejiang University, Hangzhou 310027, P. R. China}

\author{Chunmu Feng}
\affiliation{Test {\&} Analysis Center, Zhejiang University,
Hangzhou 310027, P. R. China}%

\author{Yi Xu,$^{1}$ Wen Lu,$^{1}$ Jingqin Shen,$^{1}$ Minghu Fang}
\affiliation{Department of Physics, Zhejiang University, Hangzhou
310027, P. R. China}

\author{Zhu'an Xu}
\affiliation{Department of Physics, Zhejiang University, Hangzhou
310027, P. R. China}

\date{\today}

\begin{abstract}
We report the electrical, magnetic and thermal measurements on a
layered cobalt oxyhydrate Na$_{0.31}$CoO$_{2}\cdot$1.3H$_{2}$O.
Bulk superconductivity at 4.3 K has been confirmed, however, the
measured superconducting fraction is relatively low probably due
to the sample's intrinsic two-dimensional characteristic. The
compound exhibits weak-coupled and extreme type-II
superconductivity with the average energy gap $\Delta_{a}(0)$ and
the Ginzburg-Landau parameter $\kappa$ of $\sim$ 0.50 meV and
$\sim$ 140, respectively. The normalized electronic specific heat
data in the superconducting state well fit the $T^{3}$ dependence,
suggesting point nodes for the superconducting gap structure.
\end{abstract}



\maketitle


The recent discovery of superconductivity in a two-dimensional
cobalt oxyhydrate~\cite{Takada} has been spurring new round of
intense interest in the field of superconductivity research. It
was mentioned~\cite{Takada,Badding} that the cobalt oxyhydrate
superconductor resembles the high-$T_{c}$ cuprates in the
two-dimensional (2D) MO$_{2}$ (M=Co or Cu) layers and the
existence of spin 1/2 for Co$^{4+}$ and Cu$^{2+}$ ions, though
their difference is obvious for the triangular CoO$_{2}$ sheets in
contrast with the nearly tetragonal CuO$_{2}$ planes. The fact
that the superconductivity is derived from the intercalation of
H$_{2}$O into the host Na$_{0.35}$CoO$_{2}$, which itself is not a
superconductor, suggests that strong two-dimensionality be
important for the appearance of superconductivity~\cite{Takada}.

The related theoretical work has been performed quickly, though
some basic physical property characterizations of the new
superconductors have not been reported yet. By employing the $t-J$
model on a planar triangular lattice, different kinds of
superconducting states, such as time-reversal-symmetry-breaking
$d_{x^{2}-y^{2}}+id_{xy}$ superconductivity
~\cite{Baskaran,Kumar,Wang}, and spin triplet superconductivity
~\cite{Baskaran,Tanaka} have been proposed. Based on the density
functional calculation, Singh~\cite{Singh-1} also speculates that
a triplet superconducting state may exist in this kind of
material. In a word, exotic superconductivity in the new system
seems to be a consensus for theorists. To verify the theoretical
result, therefore, the experimental investigations becomes very
crucial on this topic.

Unfortunately, the development on the experimental aspect goes
relatively slowly. One of the major reasons is that the
preparation of samples is not optimized at present. The other
reason concerns about the chemical instability of the oxyhydrate
superconductor. It was reported~\cite{Foo} that the material is
exceptionally sensitive to both temperature and humidity near
ambient conditions, which makes the experimental reproducibility
rather difficult. Consequently, only a few experimental results,
such as the magnetic properties~\cite{Sakurai} and the hydrostatic
pressure effect on $T_{c}$~\cite{Lorenz} have just been reported.
Although some unconventional magnetic properties were revealed for
the new superconductor~\cite{Sakurai}, other basic properties such
as the low-temperature specific heat have not been reported yet
for this newly-discovered superconductor. We recently succeeded in
preparing the cobalt oxyhydrate superconductor using a modified
synthetic route~\cite{Cao}. The problem of the sample's
instability was overcome to some extent by employing suitable
experimental procedure. In this Letter, we report some
superconducting and normal-state properties of this intriguing
compound.

Our samples of \ncow were prepared in four steps, briefly
described as follows. First, single-phase hexagonal
Na$_{0.74}$CoO$_{2}$ was prepared by a solid-state reaction at
1083 K in flowing oxygen with Na$_{2}$CO$_{3}$ and Co$_{3}$O$_{4}$
as the starting material. Second, partial sodium in
Na$_{0.74}$CoO$_{2}$ was deintercalated by the excessive bromine
solved in acetonitrile, similar to the treatment reported
previously ~\cite{Takada,Foo}. Third, a hydration process was
carried out by the direct reaction with distilled water. Last, the
sample was slightly dehydrated and then "annealed" under ambient
condition. Powder X-ray Diffraction (XRD) measurement indicates
that the final product is a hexagonal single phase with the cell
constants of $a$=2.820 \AA\ and $c$=19.65 \AA\ . The unit cell is
slightly stretched along the $c$-axis, compared with that of the
previous report~\cite{Takada}. This is probably due to the
difference in the Na content. By employing the Atomic absorption
spectroscopy, the atomic ratio of Na and Co was determined as 0.31
for the final product. Thermogravimetric analysis shows that the
weight loss from 293 K to 693 K is 19.8 \%, indicating that the
content of H$_{2}$O is about 1.3 per formula. Therefore, the
chemical formula of the final product is expressed as \ncow.
Details of the sample's preparation and characterizations will be
given elsewhere~\cite{Cao}.

The physical property measurements were performed at the
temperature down to 1.8 K and under the field up to 8 Tesla, on a
Quantum Design PPMS system. While measured under "zero field",
there still exists a remanent field of $\sim$ 1 Oe. The precisions
of ac magnetic susceptibility ($\chi_{ac}$) and dc susceptibility
($\chi_{dc}$) are better than $\sim 10^{-7}$ emu and $\sim
10^{-5}$ emu, respectively. The electrical resistivity ($\rho )$
was measured in a standard four-probe configuration using a
pressed sample bar. The heat capacity was measured using an
automated relaxation technique with a square piece of $\sim$ 20 mg
sample. The contribution from the addenda has been subtracted. It
is noted that the handling of the sample and the detailed
measurement procedure sometimes affect the experimental result
very much. So, we kept the same experimental condition for the
different measurements.

Figure 1(a) shows the temperature dependence of magnetic
susceptibility at low temperatures for the \ncow sample. The real
part of ac susceptibility $\chi'$ shows the onset of diamagnetism
at 4.3 K, followed by a broad superconducting transition, similar
to the original report~\cite{Takada}. The diamagnetic screening
signal at 1.9 K is 9.2 \% of the full shielding when the ac field
amplitude ($H_{ac}$) is 2 Oe, suggesting relatively low
superconducting fraction. Considering that the $\chi'$ value is
not flat down to 1.9 K, the superconducting volume fraction will
be over 10 \% under the remanent field of $\sim$ 1 Oe. The
imaginary component of the ac susceptibility shows an incomplete
dissipation peak, also suggesting that the superconducting
transition is not finished yet at 1.9 K. The dc susceptibility
under 30 Oe shows even low magnetic exclusion, which is primarily
due to the very low $H_{c1}$ value as well as the magnetic
penetration (see the result below). An irreversible temperature
can be noticed, like that observed in the high \Tc
cuprates~\cite{Muller}.

From the structural and chemical bonding points of view, the
present system should have very weak coupling between the
CoO$_{2}$ layers, resulting in a strong 2D superconductivity. It
is proposed that the relatively low superconducting fraction is
mainly due to the sample's intrinsic 2D characteristic. The
following observations are coincident with this point. First, the
superconducting transition is broad. Second, zero resistance can
never be achieved in our experiments as well as the previous
report~\cite{Takada}. Third, the diamagnetic signal is enhanced
when decreasing $H_{ac}$. Similar result was reported for a 2D
organic superconductor
(BEDT-TTF)$_{2}$Cu(NCS)$_{2}$~\cite{Manskey}. It should be pointed
out that the low superconducting fraction is \emph{not} mainly due
to the sample's instability, because our XRD experiment shows that
the sample contains no secondary phases before and \emph{after}
the magnetic susceptibility measurements.

\begin{figure}[tbp]
\includegraphics[width=8cm]{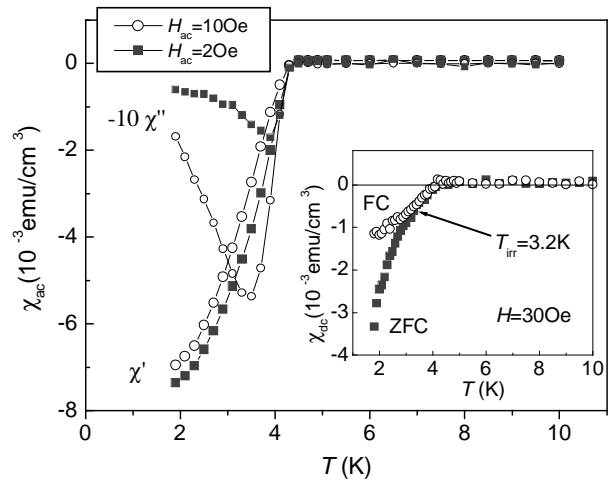}
\caption{Temperature dependence of ac magnetic susceptibility at
zero field for \ncow powdered sample. The inset shows the dc
magnetic susceptibility under the field $H$=30 Oe. $H_{ac}$,
$T_{irr}$, FC and ZFC refer to the ac field amplitude,
irreversible temperature, field cooling and zero-field cooling,
respectively.} \label{fig:chiralites}
\end{figure}

Figure 2(a) shows the magnetization loop at 1.9 K for the \ncow
sample. Narrow field hysteresis was observed, superposed with a
paramagnetic background which can be described by the Brillouin
function. The amplificatory plot using the upper-right coordinates
indicates the type-II superconductivity with $H_{c1}$ of about 10
Oe at 1.9 K. By the data fitting of $H_{c1}(T)$ using the
well-known equation: $H_{c1}(T)=H_{c1}(0)[1-(T/T_{c})^{2}]$,
$H_{c1}$(0) can be obtained as 13 Oe. The \Hc2 value is difficult
to be measured by the $M-H$ curve due to the very narrow
hysteresis. Nevertheless, by measuring the electrical resistance
at fixed temperatures, one can basically obtain the \Hc2($T$)
data, as shown in figure 2(b). \Hc2($T$) is here determined as the
point where the resistance deviates from the linearity in the
$R-H^{2}$ curves~\cite{note-1}. The slope of \Hc2 at $T_{c}$,
$dH_{c2}/dT\mid_{T_{c}}$, is obtained as $-34$ kOe/K. \Hc2(0) can
thus be estimated to be 1$\times10^{5}$ Oe, using the WHH
formula~\cite{WHH}. Then, the average Ginzburg-Landau (GL)
coherent length $\xi_{GL}$(0)=57 \AA\ can be calculated using the
formula $\xi_{GL}$(0)=$(\Phi_{0}/2\pi H_{c2}(0))^{1/2}$. On the
other hand, by solving the equation
$H_{c1}=\Phi_{0}\ln(\lambda/\xi)/4\pi\lambda^{2}$, the average
penetration depths can also be obtained: $\lambda$(0)=7900 \AA\ .
Therefore, the Ginzburg-Landau parameter $\kappa=\lambda/\xi_{GL}$
is estimated as $\sim$ 140, indicating that the cobalt oxyhydrate
is an extreme type-II superconductor, like the high-\Tc cuprates.
This conclusion has also been drawn in a very recent
report~\cite{Sakurai}, in which different method was employed to
determine the \Hc2$(T)$. It is worth while to note that, compared
with the previous result, the values of $H_{c1}$(0) and \Hc2(0) in
the present sample are remarkably smaller, which is possibly
resulted from the differences in the carrier-doping level and/or
the water content.

\begin{figure}[tbp]
\includegraphics[width=8cm]{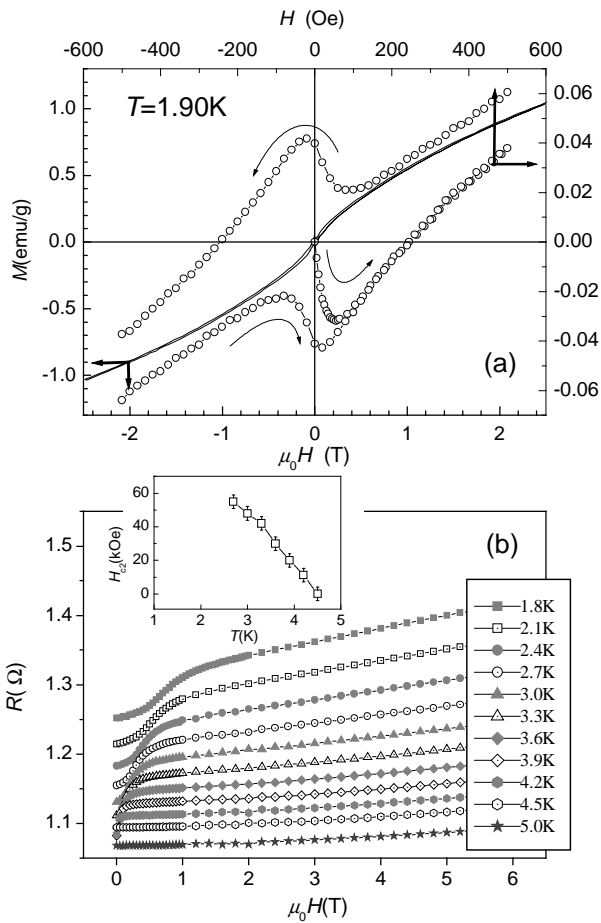}
\caption{Magnetic field dependence of magnetization (a) and
electrical resistance (b) at certain temperatures. Note that the
upper-right axes are employed for the amplificatory plot in (a).
The inset of (b) shows the temperature dependence of the upper
critical field \Hc2.}
\end{figure}

The result of low-temperature specific heat measurement is shown
in figure 3. At temperatures much below the Debye temperature
$\Theta_{D}$, and if neglected the possible magnetic contribution,
the specific heat can be expressed as the sum of electron and
phonon contributions: $C=\gamma$$T$ + $\beta$$T^{3}$, where the
coefficient $\gamma$ is generally called Sommerfeld parameter. The
phonon contribution can be separated by employing the $T^{2}$ vs
$C/T$ plot. It can be seen that good linearity is satisfied in the
temperature range of 4.5 K$<T<$11 K. We thus obtained $\gamma$ =
15.9 mJ/K$^{2}$$\cdot$mol-f.u. (f.u. denotes formula unit) and
$\beta$ = 0.235 mJ/K$^{4}$$\cdot$ mol-f.u. $\Theta_{D}$ is then
calculated to be 391 K using the formula
$\Theta_{D}=((12/5)N\pi^{4}R/\beta)^{1/3}$, where $N$=7.21 for
\ncow and $R$=8.314 J/mol$\cdot$K. The $\gamma$ value is
significantly smaller than that of the parent compound
Na$_{0.5}$CoO$_{2}$ ($\gamma\sim$ 40
mJ/K$^{2}$$\cdot$mol-Co~\cite{Ando}). Since the Sommerfeld
parameter $\gamma$ is related to the density of state (DOS) at
Fermi level, $N(E_{F})$, by the relation
$\gamma=\frac{1}{3}k_{B}^{2}\pi^{2}N(E_{F})=
\frac{1}{3}k_{B}^{2}\pi^{2}N(0)(1+\lambda_{ep})$, where $N(0)$ is
the bare, or band-structure electronic DOS at $E_{F}$,
$\lambda_{ep}$ an electron-phonon interaction
parameter~\cite{McMillan}, one can obtain that $N(E_{F})$=6.7
states/eV$\cdot$f.u. On the other hand, $\lambda_{ep}$ can be
calculated to be 0.57 using the formula
\begin{equation}
\lambda_{ep}=\frac{1.04+\mu^{*}ln(\Theta_{D}/1.45T_{c})}{(1-0.62\mu^{*})ln(\Theta_{D}/1.45T_{c})-1.04},
\label{eqlambda}
\end{equation}
where Coulomb repulsion parameter $\mu^{*}$ is assumed to be 0.13
empirically~\cite{McMillan}. Therefore, $N(0)$ is derived to be
4.3 states/eV$\cdot$f.u. We note that this value is almost
identical to the band calculation result (4.4 states/eV$\cdot$Co)
for the parent compound Na$_{0.5}$CoO$_{2}$~\cite{Singh-2}.

It is noted that the sample's magnetic susceptibility ($\sim$
2.0$\times10^{-3}$ emu/mol-f.u) is almost temperature independent
from 30 K to 300 K (not shown here). In order to obtain the Pauli
susceptibility $\chi^{Pauli}$, the $\chi(T)$ data were fitted
using the equation
$\chi=\chi_{0}+AT^{2}+C/(T-\theta)$~\cite{Sakurai}. We obtained
that $\chi_{0}$, $A$, $C$ and $\theta$ are 0.0019 emu/mol, 2.6
$\times 10^{-9}$ emu/mol$\cdot$K$^{2}$, 0.0024 emu$\cdot$K/mol and
1.7 K, respectively. The parameter $C$ gives the small effective
magnetic moment of 0.14 $\mu_{B}$. The small positive $\theta$
value suggests the existence of weak ferromagnetic correlations.
The unusually large $\chi_{0}$ value should be dominantly
contributed by $\chi^{Pauli}$, which is probably enhanced by the
Stoner-type ferromagnetic correlation. The Wilson ratio,
$R_{W}=\pi^{2}k_{B}^{2}\chi^{Pauli}/3\gamma\mu_{B}^{2}$, is
calculated to be 11, in sharp contrast with the case of heavy
fermion superconductor~\cite{Stewart}.

\begin{figure}[tbp]
\includegraphics[width=8cm]{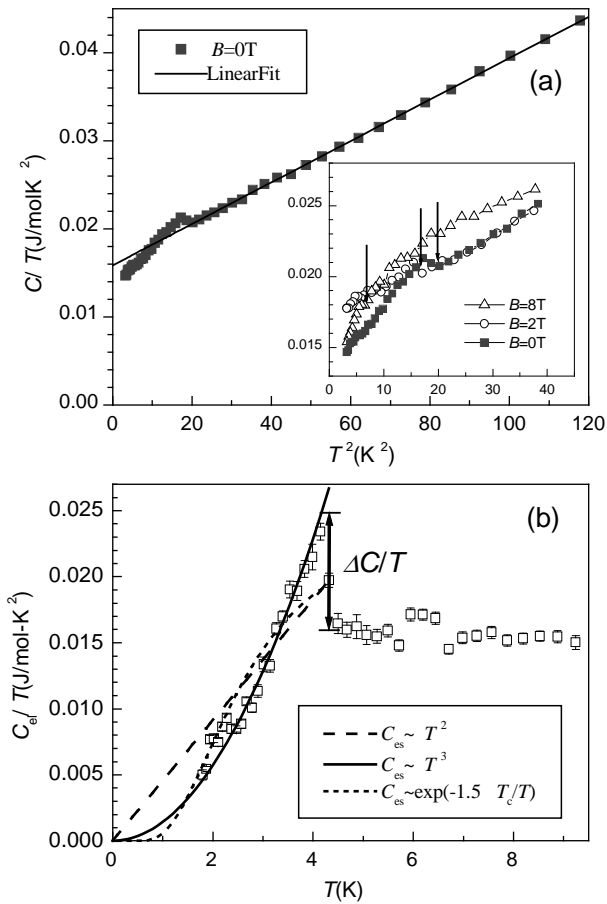}
\caption{Low-temperature specific heat result of the \ncow
superconductor. (a) plot of $C/T$ vs $T^{2}$. The arrows in the
inset point to the \Tc under different field. (b) temperature
dependence of the normalized Sommerfeld parameter. The electronic
specific heat data in the superconducting state, $C_{es}(T)$, was
fitted using different formula containing just one fitting
parameter (the coefficient).}
\end{figure}

At 4.3 K, specific heat anomalies can be seen, which is ascribed
to the superconducting transition. The specific heat jump at the
\Tc under zero field, $\Delta C_{obs}$, is 6.9 mJ/K$\cdot$mol-f.u,
further confirming the bulk superconductivity. When applying
magnetic field, both the $\Delta C_{obs}$ and \Tc decrease as
expected. It is noted that the \Tc$(H)$ values are basically
consistent with the \Hc2$(T)$ result described above.

The specific jump at \Tc, $\Delta C$, can be calculated using an
approximate formula $\Delta C=H_{c}(0)^{2}/2\pi T_{c}$, where
$H_{c}(0)$ is the thermodynamic critical field. $H_{c}(0)$ is
found to be 505 Oe by using the formula
$H_{c}(0)=H_{c2}(0)/\sqrt{2}\kappa$, where $H_{c2}(0)$ and
$\kappa$ are 1$\times10^{5}$ Oe and 140, respectively. Then,
$\Delta C$ should be 38.2 mJ/K$\cdot$mol-f.u. Therefore, the
superconducting fraction is estimated to be $\Delta C_{obs}/\Delta
C$=18.1 \%, which is basically consistent with the magnetic
susceptibility measurement result. In addition, the average
superconducting gap at zero temperature, $\Delta_{a}(0)$, can be
obtained using the relation~\cite{Goodman},
\begin{equation}
\frac{2\Delta_{a}(0)}{k_{B}T_{c}}=\frac{4\pi}{\sqrt{3}}[\frac{H_{c}(0)^{2}V_{m}}{8\pi\gamma
T_{c}^{2}}]^{1/2}. \label{eqdelta}
\end{equation}
$\Delta_{a}(0)$ is then obtained to be 0.50 meV. The value of
$2\Delta_{a}(0)/k_{B}T_{c}$ is found to be 2.71, suggesting that
the system belongs to the weak coupling limit.

A further data-analysis was carried out as follows. The lattice
specific-heat contribution, $C_{L}=\beta$$T^{3}$, was first
deducted, obtaining the electronic specific heat term:
$C_{el}=C-C_{L}$. If the superconducting fraction is $\eta$, the
electronic specific heat of the full superconductor can be
normalized as $C_{es}$=[$C_{el}-(1-\eta) \gamma T$]/$\eta$. Figure
3(b) shows the result with $\eta$=18.1 \%. The
Sommerfeld-parameter jump at the $T_{c}$, $\Delta C$/\Tc, becomes
9 mJ/K$^{2}$$\cdot$mol-f.u. So, the dimensionless parameter
$\Delta C/\gamma T$ value is about 0.57, which is remarkably lower
than the expected value 1.43 for an isotropic gap. This suggests
that the superconducting order parameter is basically not a
$s$-wave.

As we know, the temperature dependence of $C_{es}$ may give
important information on the structure of the superconducting gap.
At the temperatures far below \Tc, the temperature dependences of
$C_{es}(T)\propto$ exp$(-bT_{c}/T)$ with $b\sim$ 1.5,
$C_{es}(T)\propto T^{3}$ and $C_{es}(T)\propto T^{2}$ indicate an
isotropic BCS gap, point nodes and gap-zeroes along lines in the
superconducting gap structure, respectively~\cite{Sigrist}. Though
the extra-low temperature data is absent here due to the
experimental limitation, fitting on the present data may give a
preliminary clue. In figure 3(b), it can be seen that the $T^{3}$
dependence best fits the $C_{es}(T)$ data, suggesting point nodes
for the superconducting gap. It should be mentioned that the
$T^{3}$ dependence most favors the data in the wide range of 13 \%
$\leq\eta\leq$ 20 \% (When $\eta\leq$ 12 \%, $C_{es}$ becomes a
negative value at 1.8 K).

Based on symmetry and some preliminary experimental results,
Tanaka and Hu~\cite{Tanaka} proposed spin triplet
superconductivity in the cobalt oxyhydrate. The $p$-wave
superconductivity was also suggested by Baskaran~\cite{Baskaran}
for the higher doping level. Owing to the ferromagnetic
correlation in the normal state, as stated above, spin-triplet
$p$-wave pairing is very probable. Considered the point nodes for
the superconducting gap, therefore, the gap function will be
$\Delta(\textbf{k})$=$\hat{\textbf{x}}k_{x}$+$\hat{\textbf{y}}k_{y}$,
which shows the difference from the conclusion in the strontium
ruthenate superconductor~\cite{Maeno}. Obviously, further
experiments such as NMR, neutron scattering, and $\mu$ SR will be
needed to make clearer picture for the symmetry of the
superconducting order parameters.

\begin{acknowledgments}
The authors are indebted to H. H. Wen for the earlier discussion.
This work was supported by NSFC with the Grant No. 10104012 and
NSFC 10225417. Cao, Fang and Xu also acknowledge the partial
support (Project No. NKBRSF-G1999064602) from the Ministry of
Science and Technology of China .
\end{acknowledgments}

\end{document}